\documentclass
[aps,twocolumn,prl,showpacs,floatfix,superscriptaddress]{revtex4-1}%
\usepackage{graphicx}
\usepackage{amsmath}
\usepackage{amsfonts}
\usepackage[normalem]{ulem}
\usepackage{amssymb,color}
\usepackage{amssymb}%
\providecommand{\U}[1]{\protect\rule{.1in}{.1in}}
\begin{document}
\title{Localized itinerant electrons and unique magnetic properties of SrRu$_{2}%
$O$_{6}$}
\author{S. Streltsov}
\affiliation{M.N. Miheev Institute of Metal Physics of Ural Branch of Russian Academy of
Sciences, 620137, Ekaterinburg, Russia}
\affiliation{Ural Federal University, Mira St. 19, 620002 Ekaterinburg, Russia}
\author{I. I. Mazin}
\affiliation{Code 6393, Naval Research Laboratory, Washington, DC 20375, USA}
\author{K. Foyevtsova}
\affiliation{Quantum Matter Institute, University of British Columbia, Vancouver, British
Columbia V6T 1Z4, Canada}
\date{\today }

\begin{abstract}
SrRu$_{2}$O$_{6}$ has unique magnetic properties. It is characterized by a
very high N\'eel temperature, despite its quasi-two-dimensional structure, and
has a magnetic moment more than twice reduced compared to the formal ionic
count. First principles calculations show that only an ideal N\'eel ordering in
the Ru plane is possible, with no other metastable magnetic solutions, and, highly
unusually, yield dielectric gaps for both antiferromagnetic and nonmagnetic
states. We demonstrate that this strange behavior is the result of the formation of
very specific electronic objects, recently suggested for a geometrically
similar Na$_{2}$IrO$_{3}$ compound, whereby each electron is well localized on
a particular Ru$_{6}$ hexagon, and completely delocalized over the
corresponding six Ru sites, thus making the compound $both$ strongly localized
and highly itinerant.

\end{abstract}
\maketitle

The recently discovered\cite{Hiley} SrRu$_{2}$O$_{6}$ has attracted
considerable attention because, despite being a very 2D material, it shows an
exceptionally high N\'{e}el temperature of $\sim$560 K \cite{Arita,Walton}. As we
will argue in this paper, this is by far not the only, and maybe not even the
most intriguing property of this material. Ru$^{5+}$ has a half-filled
$t_{2g}$ electronic shell, and exhibits insulating behavior. Naturally, it was
interpreted as a Slater insulator (maybe Mott-enhanced), with Ru in the high
spin state, $S=3/2.$ However, the experimentally measured ordered magnetic
moment is only $M=1.3-1.4$ $\mu_{B}$ \cite{Walton,Arita}, 2.3 times smaller than expected 
for $S=3/2$ $(M=3$ $\mu_{B})$.
This was ascribed to hybridization with oxygen\cite{Singh,Arita,Walton}, but
it should be noted that such strong suppression of magnetic moment in a good
insulator is unheard of. Even in the metallic SrRuO$_{3}$ the hybridization
suppresses the total magnetic moment of Ru$^{4+}$ only from 2 to 1.7 $\mu
_{B},$ and in Sr$_{2}$YRuO$_{6}$ Ru$^{5+}$ has essentially exactly 3 $\mu
_{B},$ with basically the same Ru-O distances as in SrRu$_{2}$O$_{6}$
\cite{MazinSingh}. Hiley \textit{et al} \cite{Walton} mention the case of
Li$_{3}$RuO$_{4}$ \cite{Manuel}, where a suppression down to $M=2.0$ $\mu_{B}$
was reported for the same oxidation state, which is, however, still twice
smaller a reduction compared to SrRu$_{2}$O$_{6},$ and the material might
actually be a metal (no transport data have been published).

Electronic structure calculations~\cite{Singh, Arita} so far have not resolved
the mystery, but have only added to the confusion. It was found that only the
ideal N\'{e}el state can be stabilized in the calculations, even though ions
with $S=3/2$ are usually very stable and while they disorder with temperature,
never lose their magnetic moment completely. At the same time the moment found in
the calculations matches the experimentally measured one within 8\%,
suggesting that the role of Coulomb correlations beyond the standard density
functional theory (DFT) is negligible\cite{note1}. The instability of
the ferromagnetic (FM) state was traced down to the presence of a dielectric
gap in nonmagnetic calculations\cite{Singh}, but that essentially translates
one mystery into another: why does a highly symmetric Ru sublattice, with no
dimerization or clusterization, with a half-filled $t_{2g}$ band, show a
sizeable nonmagnetic gap? Singh mentions\cite{Singh} that the gap is allowed
by symmetry, since the unit cell includes two Ru atoms that can, in principle,
form a bonding and an antibonding band, but does not elaborate about how a
structure with each Ru having three equivalent bonds manages to develop a
bonding-antibonding splitting.

Similarly, it was pointed out that, even though SrRu$_{2}$O$_{6}$
is extremely 2D magnetically, there is still some residual interlayer coupling, $J_{\perp
}M^2\approx1.5$ meV, as well as a single-ion magnetic anisotropy,
estimated to be $\approx1.4$ meV/Ru \cite{Singh}. It was suggested that the
anisotropy \cite{Singh} or interlayer coupling \cite{Arita} are responsible
for the large $T_{N},$ implying that the (unknown) mean field transition
temperature is extremely high. Tian $et$ $al$\cite{Arita} attempted to
describe this system by a three nearest neighbor Heisenberg model with
parameters derived within the perturbation theory in the limit of a Hubbard
$U$ much larger than the hopping, $U\gg t$. However, the fact that ferromagnetic
arrangement is completely unstable {(in fact, as we show below, no parallel
nearest neighbor moments are stable)}, indicates that the system is strongly
non-Heisenberg, casting very strong doubt on the relevance of such models.  
Additionally, the 
fact that the system is very weakly correlated makes such a perturbation
theory unphysical. 
Hiley \textit{et al} \cite{Walton} used a hybrid functional
that overestimates the equilibrium magnetic moment and thus the 
exchange parameters,\cite{noteH} as well as yields a very large band gap 
of 2.15 eV, totally inconsistent with the observed weak temperature dependence 
of the resistivity. 

\begin{figure}[t]
\centering
\includegraphics[clip=false,width=0.5\textwidth]{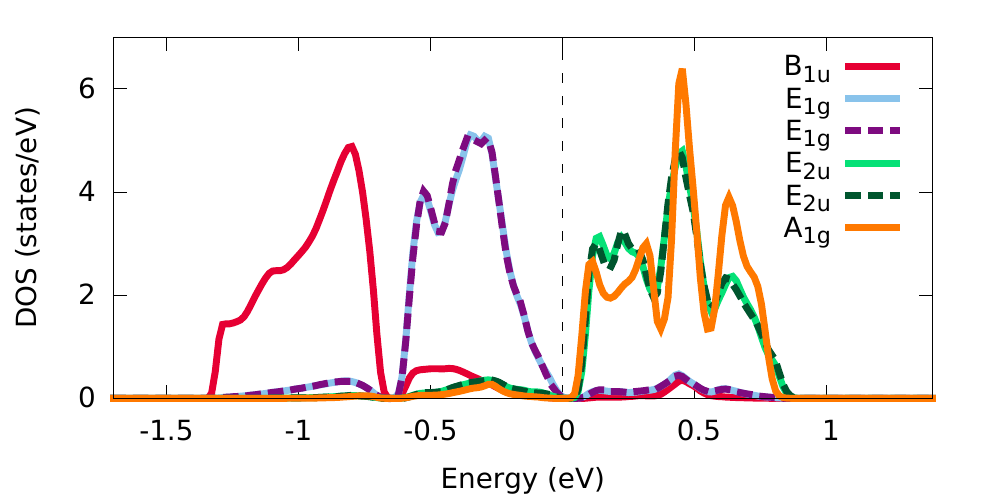}\caption{ Density of
states (DOS) projected on molecular orbitals of different symmetries in
nonmagnetic GGA calculations (WIEN2k results). The Fermi energy is set to zero.}
\label{DOS}
\end{figure}

An explanation of all these oddities can be consistently found in the
so-called molecular orbitals (MO) picture, which was first brought up in
connection to Na$_{2}$IrO$_{3}$\cite{Kateryna} and later found also in
RuCl$_{3}$ \cite{RC}. Basically, this picture is based on the idea that for
ideal 90$^{\circ}$ Ru-O-Ru bond angles (the actual angles are 101$^{\circ
}$) the O-assisted Ru-Ru hopping is only allowed for one particular pair of
the $t_{2g}$ orbitals for each hexagonal bond, denoted $t_{1}^{\prime}$ in
Ref. \cite{Kateryna}. If all other hoppings are neglected, it leads to a
curious situation where every electronic state is fully delocalized over a
particular hexagon, but never leaves this hexagon. One can say that the
electrons are fully localized (form nondispersive levels) and fully
delocalized (each state is an equal weight combination of six orbitals
belonging to six different sites). If a direct overlap of the $t_{2g}$
orbitals (which always exists in the common edge geometry) is included, as
well as deviations of the angle from 90$^{\circ}$, two more hoppings emerge:
one between the same orbitals on the neighboring sites, $t_{1},$ and the other an
O-assisted second neighbor hopping between unlike orbitals, $t_{2}^{\prime}.$
As long as $t_{1}^{\prime}$ is dominant, the MO model still applies, and can
be readily solved. The solution entails six bands, $A_{1g}$, $E_{2u}$,
$E_{1g}$ and $B_{1u}$ (the $E$ bands being double degenerate in each spin
channel), whose dispersion is controlled by $t_{1},$ and whose centers are located
at $2(t_{1} ^{\prime}+t_{2}^{\prime}),$ $(t_{1}^{\prime}-t_{2}^{\prime}),$
$-(t_{1} ^{\prime}+t_{2}^{\prime})$ and $-2(t_{1}^{\prime}-t_{2}^{\prime}),$
respectively. In Na$_{2}$IrO$_{3}$ $t_{1}^{\prime}\approx-3t_{2}^{\prime},$ so
that the $A_{1g}$ and $E_{2u}$ practically merge. This accidental degeneracy
also leads to much stronger spin-orbit effects than would have been possible
had the MO bands remained well separated, and to considerable destruction of the
MO picture in the relativistic case. On the other hand, the hopping parameters for
SrRu$_{2}$O$_{6},$ as calculated in Ref. \cite{Wang}, are similar to those in
Na$_{2}$IrO$_{3},$ in the sense that again $t_{1}^{\prime}=300$ meV is by far
the largest hopping, and the only other sizeable hoppings are $t_{1}=160$ meV
and $-t_{2}^{\prime}\approx100-110$ meV. Note that here $|t_{2}^{\prime}|$ is
again about $1/3$ of $t_{1}^{\prime}$. Thus, the $A_{1g}$ and $E_{2u}$ bands
merge, while $E_{1g}$ and $B_{1u}$ remain separated, as one can see in Fig.
\ref{DOS}. Projecting the density of states onto MOs, we observe that the
predicted characters are very well reproduced. The distance between the centers of
the $E_{2u}$ and $E_{1g}$ bands is about 0.8 eV, and their width is about 0.6-0.7
eV, thus providing for a small gap of $\approx50$ meV.

It is instructive to compare SrRu$_{2}$O$_{6}$ with Na$_{2}$IrO$_{3}$ and with
Li$_{2}$RuO$_{3}.$ All these compounds share the same crystallographic motif,
but feature a different number of $d$ electrons: 5, 4 or 3. In the iridate, a
single hole in the upper $A_{1g}$ singlet is prone to both strong correlations
and, due to near degeneracy between $A_{1g}$ and $E_{2u},$ to spin-orbit
interaction. As a result, as one increases the spin-orbit coupling, the
$A_{1g}$ singlet is gradually transformed into the $j_{eff}=1/2$
singlet~\cite{Kateryna}. Either way, a half-filled singlet triggers Mott
physics even if the Hubbard $U$ is small. This transformation controls most of
the interesting physics in this compound. Li$_{2}$RuO$_{3}$ has two $d$ holes,
providing it with an opportunity to form strongly bound covalent dimers. This
is exactly what happens, and the MO on the hexagons transforms to an MO on the
Ru dimers resulting in the spin singlet ground state~\cite{Kimber}. Neither
Mott nor spin-orbit physics is relevant on the background of the strong
covalent bonding in dimers. Finally, SrRu$_{2}$O$_{6}$ has the six MO bands
half-filled, and the gap is formed between the lower and the upper MO triads.
Similar to Li$_{2}$RuO$_{3},$ both Mott and spin-orbit effects are of minor
importance, and the gap structure inherent to the MO picture gives rise to
unique magnetic properties.

Let us now turn to the energetics of the material. First, we have confirmed,
using the WIEN2k package \cite{wien2k,sup}, the numbers published by
Singh\cite{Singh} regarding the interplanar coupling, single-site anisotropy
and Ru magnetic moment. We also confirmed that the ferromagnetic structure
cannot be stabilized. Moreover, the so-called stripy and zigzag magnetic
patterns\cite{Kateryna}, where one or two out of three bonds are ferromagnetic,
 and the net moment is
zero, cannot be stabilized.  This indicates
that besides the obvious influence of the nonmagnetic gap there are other
factors strongly disfavoring ferromagnetic bonds. In fact, given that the gap
value is ten times smaller than Ru Stoner factor\cite{MazinSingh}, and the
calculated magnetic moment in the N\'{e}el state is $\sim$1.3 $\mu_{B},$ it is
surprising that the ferromagnetic bonds do not stabilize with a finite moment.
\begin{figure}[t]
\centering
\includegraphics[clip=false,width=0.5\textwidth]{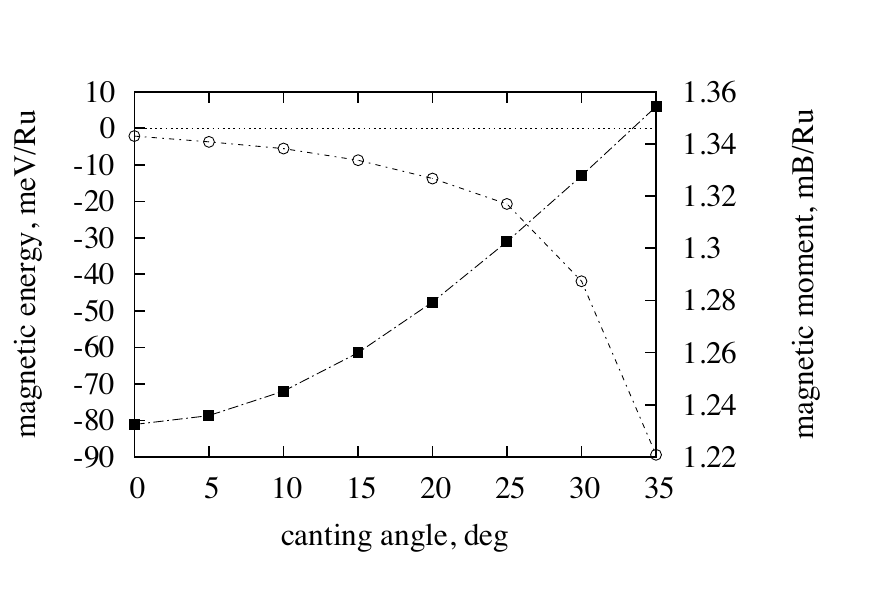}\caption{Magnetic
energy (squares) and magnetic moments (circles) as a function of the canting
angle of spins, starting from the N\'{e}el antiferromagnetic structure.
Results are from VASP calculations.}
\label{angle}
\end{figure}

In order to gain more insight into the problem, we turned to the VASP code
\cite{vasp,sup}, which is faster and has the capability to restrict magnetic
moments to a certain direction, or to both a direction and a magnitude (we
confirmed that the energies of collinear magnetic states agree with those 
found in WIEN2k). First, we computed the total energy for a canted
antiferromagnet (AFM), restricting the angle with the $z$ axis to be $\pm\phi$
for the two Ru's in the cell. The results are shown in Fig. \ref{angle}. Note
that for the largest canting angle we were able to converge, 35$^{\circ}$, the
energy of the magnetic state is already higher than that of the nonmagnetic
one. Also note how soft the magnetic moments are: despite the sizeable equilibrium
moment, the energy cost of total suppression of magnetism is less than 80 meV,
only 50\% larger than the transition temperature. This is, again, an
indication of the great role of itinerancy, and specifically, delocalization over
Ru$_{6}$ hexagons.

Interestingly, suppression of magnetism with canting cannot be described by a
naive combination of a local Hamiltonian for itinerant magnets\cite{Moriya},
$E=\sum\limits_{i\geq0}a_{i}M^{2i}$, where $M$ is the magnetization, with a
Heisenberg term.  While the total energies at a fixed canting angle
$\phi\lesssim35{{}^{\circ}}$ can be very well described by this Hamiltonian
with just three terms, $E(M)-E(0)=a_{1}M^{2}+a_{2}M^{4}+a_{3}M^{6},$ not only
does the first coefficient show appreciable angular dependence (as in the
Heisenberg model), but also the second, and, to a lesser degree, the third.
Instead, a good fit could be obtained with the following formula:
\begin{align}
E &  =-81.3M^{2}+16.9M^{4}+2.0M^{6}+359.2M^{2}\sin^{2}\left(  \phi\right)
\nonumber\\
&  -165.8M^{4}\sin^{2}\left(  \phi\right)  +27.6M^{6}\sin^{2}\left(
\phi\right)  ,\label{E}%
\end{align}
in meV/Ru. Note that the angle between the moments is $\theta=\pi-2\phi,$ and
that there are 1.5 times more bonds than sites. Thus, the proposed Hamiltonian
looks as follows:
\begin{align}
H &  ={\sum\limits_{sites}}\{98.3M^{2}-66.0M^{4}+15.8M^{6}\}\label{H} \nonumber \\
&  +{\sum_{\substack{n.n. \\ bonds}}}\{179.6  (\mathbf{M\cdot M}^{\prime})
-82.9 |\mathbf{M}| | \mathbf{M'}|  (\mathbf{M\cdot M}^{\prime})\\
&+13.8 |\mathbf{M}|^2 | \mathbf{M'}|^2 (\mathbf{M\cdot M}^{\prime})\}.\nonumber
\end{align}
The Heisenberg term is extremely strong $(JM^{2} = \partial H/\partial \cos \left( \theta \right) 
\approx 1600$ K), and, without
it, local magnetic moments fail to form.

To this Hamiltonian one needs to add a small interlayer term $\sum J_{\perp
}\mathbf{M}_{i}\mathbf{\cdot M}_{i^{\prime}},$ where $i$ and $i^{\prime}$
belong to the neighboring planes, and the magnetic anisotropy $\sum DM_{z}%
^{2},$ where $J_{\perp} \approx0.9$ meV, and $D \approx0.8$ meV.

In principle, at this point one would need to perform a Monte Carlo simulation
using this Hamiltonian and determine the transition temperature. However, it
is notoriously difficult to distinguish a Kosterlitz-Thouless phase in a
quasi-2D system from the true long range order, so that one should be very skeptical
of any Monte Carlo simulation that claims to establish a N\'{e}el temperature,
$T_{N},$ without first showing that in the isotropic 2D limit $T_{N}$ truly
vanishes. The softness of the moment, expressed $via$ Eq. \ref{H},
additionally complicates the simulation. We leave this daunting task to more
experienced Monte Carlo simulators, but mention that the numbers that we have
deduced are in the right ballpark. For instance, Costa and Pires
showed\cite{Costa} that for the square lattice $T_{N}/T_{MF}\approx
0.8(D/J)^{0.2}.$ For three neighbors, the mean field transition temperature
$T_{MF}\approx JM^{2}\approx 1600$ K, which together with $DM^{2}\approx1.4$
meV results in $T_{N}\sim 500$ K. On the other hand, for the cubic quasi-2D
model with $J_{\perp}\neq0,$ $D=0,$ Yasuda $et$ $al$\cite{Yasuda} found that
$T_{N}\approx4.27JM^{2}/[3.12+\log(J/J_{\perp})],$ which for our parameters
translates into 900 K. Thus, we conclude that (a) the Mermin-Wagner theorem
is mainly lifted $via$ the interplanar coupling\cite{Arita}, and not $via$ the
single site anisotropy\cite{Singh}, and (b) the softness of the magnetic
moment, $i.e.,$ longitudinal fluctuations, plays an important role,
suppressing $T_{N}$ by up to a factor of two.

\begin{figure}[t]
\centering
\includegraphics[clip=false,width=0.5\textwidth]{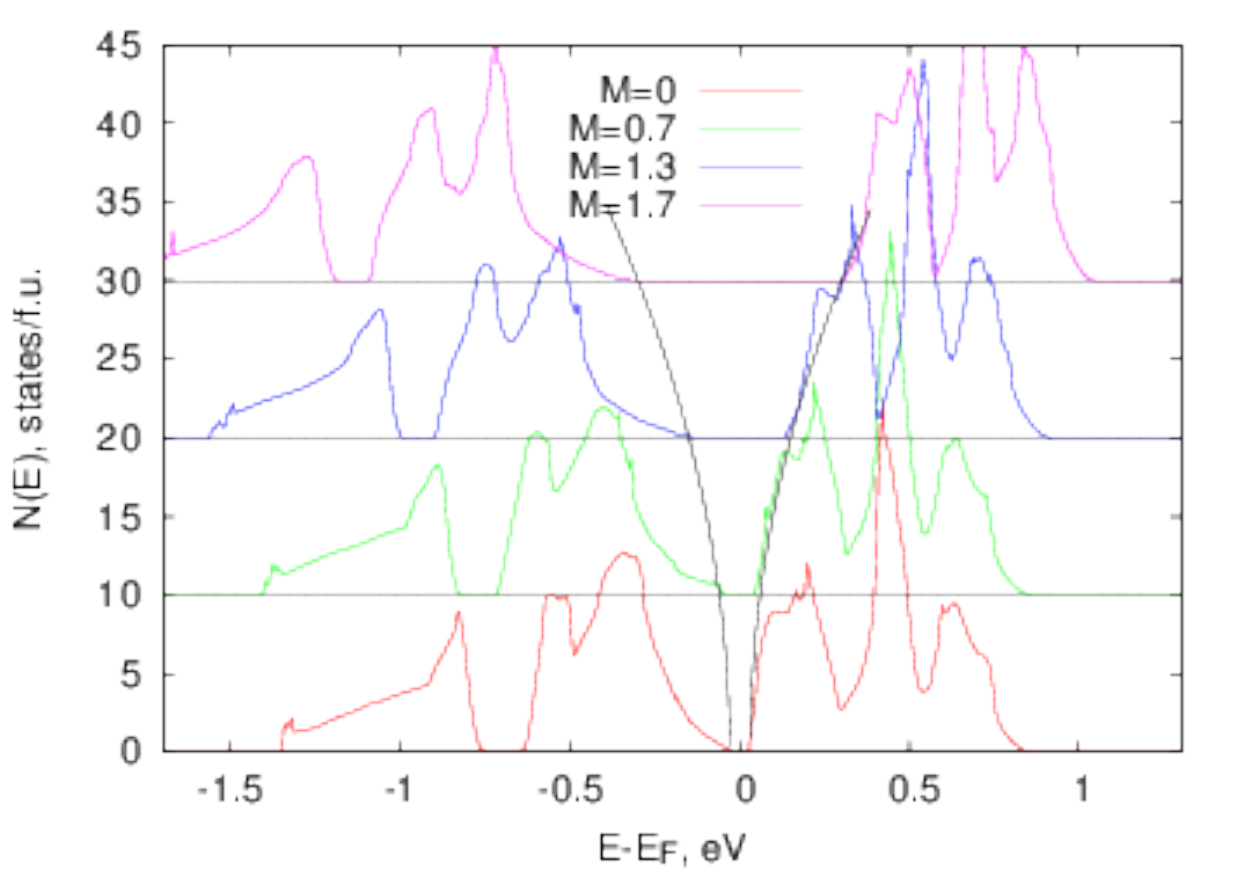}\caption{Total
density of states (DOS) calculated for several values of Ru moments,
$M$, in the fixed-spin-moment procedure for the N\'{e}el AFM. The black
lines illustrate the fact that he band gap is approximately quadratic in $M$.
 Results of the LAPW calculations.}%
\label{FSM}%
\end{figure}

Let us now discuss how and why MOs support a N\'{e}el antiferromagnetism in
SrRu$_{2}$O$_{6}$. In the nonmagnetic state, the three lower MO bands,
$B_{1u}$ and $E_{1g},$ are fully occupied. Imposing uniform spin polarization
does not change the occupancy of these states, unless the induced exchange
splitting is larger than the gap, and this is why the ferromagnetic order is
unstable. On the contrary, imposing the staggered magnetic field of $\pm\Delta$
does not break the MO band structure, but rather increases the gap between
$E_{1g}$ and $E_{2u}$ (in the lowest order in $\Delta,$ by $\Delta^{2}%
/t_{1}^{\prime})$. In the same order, we can calculate the change of the
occupancies and find that the spin-up sites acquire magnetization of
5$\Delta/2t_{1}^{\prime}$ $\mu_{B},$ and the spin-down sites $-5\Delta
/2t_{1}^{\prime}$ $\mu_{B}.$ The signs are consistent with the assumed signs
of $\Delta,$ which tells us that with sufficiently large Hund's rule coupling
the system will become unstable against such a staggered magnetization (but
will resist any ferromagnetic component); of course, quantitative analysis is
impossible on this level of simplification. Obviously,
the equilibrium moment can be anything between 0 and 3 $\mu_{B}.$ It is not
\textquotedblleft suppressed\textquotedblright\ from the putative $S=3/2$ state,
but is set by the interplay between the Hund's rule coupling on Ru and the
details of the density of states of MOs. A corollary from the above arguments
is that the dielectric gap depends quadratically on the Ru moment;
Fig.~\ref{FSM} illustrates that this is indeed the case, to a reasonable
accuracy.

\begin{figure}[t]
\centering
\includegraphics[clip=false,width=0.5\textwidth]{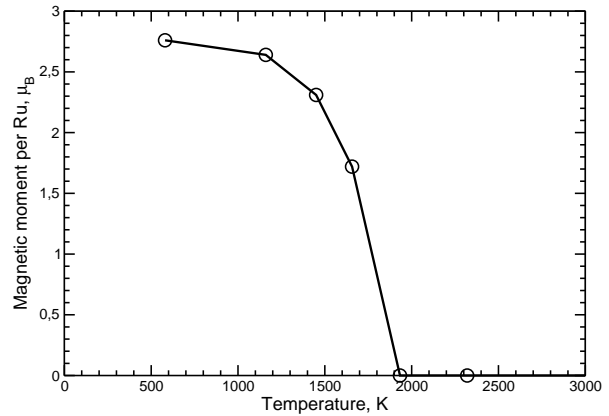}\caption{Magnetic
moment as calculated in the LDA+DMFT approach. The continues-time quantum
Monte-Carlo (CT-QMC) solver \cite{Werner06} was used in these calculations.}%
\label{DMFT}%
\end{figure}

Let us emphasize, that the molecular orbitals are not just another way to
describe the electronic structure of SrRu$_{2}$O$_{6}$, but have profound
physical meaning. It is instructive to compare it with another recently
investigated high-$T_{N}$ material, SrTcO$_{3}$, where the transition metal
also has a $4d^{3}$ configuration and $S=3/2$. It was argued \cite{Georges}
that $T_{N}$ is so high because SrTcO$_{3}$ is in an intermediate regime
between itinerancy and localization, which is optimal for magnetic
interactions. Indeed, LDA+DMFT calculations, well suited to this regime, have
been performed by Mravlje et al. \cite{Georges}, who found $T_{N}\approx2200$
K. The experimental number is about 1100 K. To compare this result with
SrRu$_{2}$O$_{6},$ we have also performed LDA+DMFT calculations with the AMULET code\cite{AMULET}, using an
effective Hamiltonian constructed for Ru $t_{2g}$ orbitals and interaction
parameters $U=2.7$ and $J=0.3$ eV as calculated in Ref.~\cite{Arita}
(parameters for Tc are very similar). The corresponding temperature dependence
of the magnetic moment is shown in Fig. \ref{DMFT}. Not surprisingly, we found
about the same N\'{e}el temperature (2000 K) as Mravlje et al. \cite{Georges},
and an even larger magnetic moment ($M\approx$2.7 $vs.$ 2.5 $\mu_{B}).$ The
difference, however, is that experimentally in SrRu$_{2}$O$_{6}$ both $T_{N}$
\textit{and }$M$ are about twice smaller than in SrTcO$_{3}.$ Mravlje
\textit{et al.} ascribed their overestimation of $T_{N}$ to nonlocal
fluctuations, missing in the DMFT, but observed no reduction in the ordered
moment at all, while in our case the reduction in $both$ $T_{N}$ $and$ $M^{2}$
is of the same order, about a factor of 4. This clearly indicates that there
is a fundamental difference between the two compounds, going much beyond just
the difference in dimensionality, which is related to the presence of MOs in
one and their absence in the other compound.
A proper account of the molecular orbitals within DMFT can only be done in the
cluster extension of this method \cite{Biroli2002}, which could shed more
light on this compound. 

\begin{figure}[t]
\centering
\includegraphics[clip=false,width=0.5\textwidth]{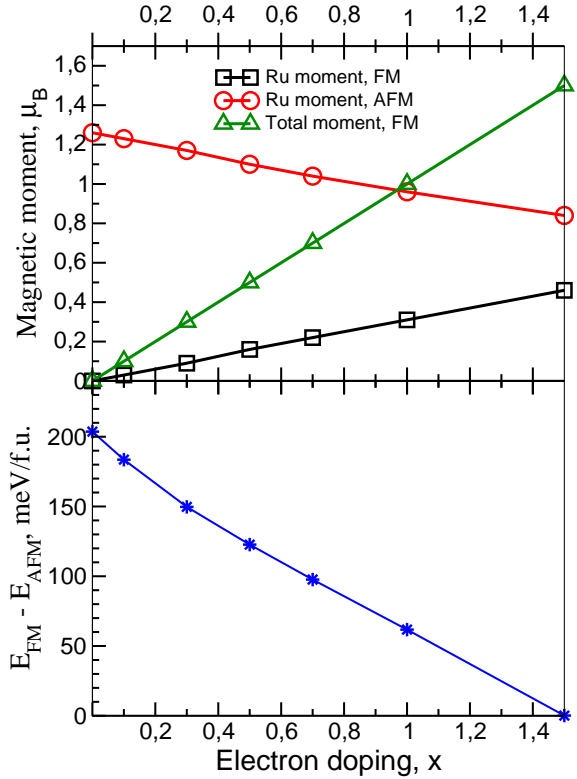}\caption{Electron
doping dependence of magnetic moments (per Ru and total) and total energy
difference between FM and N\'{e}el AFM states on the electron doping. }%
\label{doping}%
\end{figure}

Another interesting question that arises in connection with this material is what
would happen if it were doped with, for instance, a rare earth element. To address
this scenario, we simulated doping by adding electrons to the system (with a
compensating constant background). The energy difference between the FM and N\'{e}el AFM
states decreases upon electron doping, as seen from Fig.~\ref{doping}. The FM
configuration immediately becomes metastable, whereby all doped electrons go
into one spin subband, rendering the material half-metallic. The ground state
remains antiferromagnetic, but its energy advantage is gradually decreasing.
Thus, one expects that the critical angle $\phi$ (which was $\sim$35$^{\circ}$
in undopped case) will grow with doping, and the Hamiltonian \eqref{H} will be
correspondingly modified; this may result in a rapid change of magnetic
properties with doping, which deserves further theoretical and experimental investigation.

To summarize, we have found that:

(i) The electronic structure of SrRu$_{2}$O$_{6}$ is dominated by molecular
orbitals. Each electron is, to a good approximation, localized on a particular
Ru$_{6}$ hexagon, and completely delocalized over the corresponding six Ru sites.

(ii) This structure sports an excitation gap that prevents formation of
ferromagnetic bonds, but is consistent with nearest neighbor
antiferromagnetism. The corresponding magnetic interactions cannot be mapped
onto a localized spin model, be it Heisenberg or biquadratic Hamiltonian with
arbitrary long range. Neither can it be described as purely itinerant magnetism,
but features interesting elements of both. This duality reflects the dual
character of the electronic structure, where electrons are simultaneously
completely delocalized and strongly localized on the Ru hexagons. A corollary
is that any deviation from the collinear N\'{e}el order is severely punished
by kinetic energy, which, in turn, provides for the anomalously large
transition temperature.

(iii) The gaps in the non-magnetic and antiferromagnetic states have the same
nature, and one is continuously transformed into the other as the
magnetization increases. On the contrary, the ionic picture assigning the
moment of 3 $\mu_{B}$ to each Ru and associating the gap in magnetic states
with spin-up/spin-down splitting is qualitatively incorrect. The observed and
calculated magnetic moment of 1.3 $\mu_{B}$ is a manifestation of the
molecular orbital nature of electronic states, and should not be viewed as a
spin $S=3/2$ reduced by hybridization.

(iii) The magnetic properties of doped SrRu$_{2}$O$_{6}$ (e.g. by Na or La)
are expected to be very different from the stoichiometric case. One may
anticipate interesting and very different physics emerging, which can be a
subject of forthcoming research.

\textit{Acknowledgments}. S.S. and I.M. are grateful to R. Valenti and
University of Frankfurt (where this work was started) for the hospitality and
to A. Ruban, S. Khmelevskii, A. Poteryaev, D. Khomskii and K. Belashchenko for
useful discussions. This work was supported by Civil Research and Development
Foundation via program FSCX-14-61025-0б the Russian Foundation of Basic
Research via Grant No. 13-02-00374, and FASO (theme “Electron” No. 01201463326). 
I.M. is supported by ONR through the NRL
basic research program.


\begin{thebibliography}{99}
\bibitem {noteH}These authors report Monte-Carlo simulations for the
calculated set of parameters, and claim a good agreement with the experiment;
however, their nearest-neighbor only simulations yielded an unphysical result
of $T_{N}=JM^{2},$ which is the mean field temperature, rather than $T_{N}=0,$
as required by the Mermin-Wagner theorem, indicating that thermal fluctuations
have not been properly accounted for in these simulations.
\end{thebibliography}
\end{document}